\documentstyle[12pt]{article}
\def\be{\begin{equation}}
\def\ee{\end{equation}}
\def\bea{\begin{eqnarray}}
\def\eea{\end{eqnarray}}

\begin{document}
\begin{center}
{\Large \bf Moduli Thermalization and Finite Temperature Effects in ``Big" Divisor Large Volume $D3/D7$ Swiss-Cheese Compactification}
\vskip 0.1in {Pramod Shukla\footnote{email: pmathdph@iitr.ernet.in, pramodmaths@gmail.com}\\
Department of Physics, Indian Institute of Technology,
Roorkee - 247 667, Uttaranchal, India}
\end{center}
\thispagestyle{empty}

\begin{abstract}
In the context of Type IIB compactified on a large volume Swiss-Cheese orientifold in the presence of a mobile space-time filling $D3$-brane and stacks of fluxed $D7$-branes wrapping the ``big" divisor $\Sigma_B$ of a Swiss-Cheese Calabi Yau in ${\bf WCP}^4[1,1,1,6,9]$, we explore various implications of moduli dynamics and discuss their couplings and decay into MSSM (-like) matter fields early in the history of universe to reach thermal equilibrium. Like finite temperature effects in O'KKLT, we  observe that the local minimum of zero-temperature effective scalar potential is stable against any finite temperature corrections (up to two-loops) in large volume scenarios as well. Also we  find that moduli are heavy enough to avoid any cosmological moduli problem.
\end{abstract}

%

\section{Introduction}

Moduli stabilization and realizing de-Sitters solutions in string framework, being a prerequisite for realistic model building in string cosmology,  has been of quite an interest since long and a lot of progress have been made (see \cite{KKLT,Balaetal2,dSetal} and references therein). The plethora of moduli appearing after various compactification processes have very exciting role in real world physics, as these could be possible candidates for inflaton, dark energy scalar field as well as a possible quintessence field and may other interesting physical implications theirof, developing a completely new and active subarea ``moduli cosmology" which  has attracted some deep attention in recent few years. Some moduli with large life time(s) have also been realized  to be a possible  dark matter candidate \cite{AstroLVS}. Inspite of their various useful shadows in explaining real world, on the black side of the picture moduli have been troubling the researchers from the very beginning of there appearance/origin starting from moduli stabilization and also cause several cosmological problems yet to be resolved in a satisfactory way and hence the ``moduli dynamics" (including the study of their various possible couplings to gravitational and MSSM-like spectrum) needs to be investigated in a proper analytic as well as numeric manner. Further the stability of realized de-Sitter solutions against various known/unknown corrections is an important issue to investigate among various others challenging tasks like embedding inflation scenarios and explain structure formations, dark energy/matter via realistic model building. Hidden sector plays a very crucial role on real world physics through various aspects with moduli, like affecting vacuum structure in de-Sitter realizations, structure of supersymmetry breaking etc. while studying the stability of de-Sitters as the various hidden sector moduli and visible sector matter couplings are dictated by the internal compactification geometries.

There has been a famous issue of ``cosmological moduli problem" which arises if the moduli  carry very large amount of energy spoiling the successes of big bang nucleosynthesis \cite{CMP1,CMP2,CMP3}. This problem can be understood as follows: when a modulus is displaced from the minimum of the effective potential, it coherently oscillates around its minimum with the decrease in the energy of these oscillations like $a^{-3}$ whereas radiation decreases as $a^{-4}$, where ``$a$" is the scale/expansion factor of the universe appearing in FRW metric. Therefore the moduli may soon dominate the total energy density of the Universe and can cause an overclose of the Universe or at some point, there arises a possibility of spoiling the success of standard big bang nucleosynthesis. In order to avoid the moduli problem, some huge entropy production to dilute moduli density is required and the most successful mechanism for entropy production is "thermal inflation" \cite{Thermalinflation1}, however this might also dilute primordial baryon asymmetry of the universe. This problem of baryon asymmetry has been settled in \cite{Thermalinflation2} in the context of thermal inflation itself. Several attempts with resolutions to cosmological moduli problem have been proposed (see \cite{CMP1,CMP2,CMP3,Thermalinflation1,Thermalinflation2,LindeCMP} and references therein). One possible candidate for resolving this problem is the heavy moduli scenarios which has a situation of moduli being heavier enough than sparticles (in the visible sector spectrum) implying moduli decay reheating the universe to temperatures above nucleosynthesis temperature and hence the problem getting settled. However the production of dark matter and gravitinos are possibly arising obstacles to be addresses in such scenarios and some recent progress can be found in \cite{DMCMP} (and the references therein). Another resolution has been proposed in the context of hot early Universe with arguments that the moduli oscillations in hot early universe could be damped enough in the vacuum such that the moduli would adiabatically proceed towards the minimum of the effective potential without creating any problem for cosmology \cite{hotUCMP}. Due to small coupling constants in gravitational-strength interactions (which are usually Planck suppressed) involving the moduli, it has been observed that usually these are either stable or decay late in the history of universe. This way moduli also play extremely crucial role in the post-inflationary scenarios like realizing non-Gaussianities and signatures of gravity waves via estimating finite/detectable tensor-to-scalar ratios in string inspired models.

The investigation of finite temperature effects, which could modify the scalar potential
washing out the local physical minimum due to the $T^4$ contribution to the scalar potential which comes from the coupling of the modulus to a thermal matter bath, has been done in several models \cite{AstroLVS,Buchmuller1,Buchmuller2,FirstFT,CicoliFTLVS}. In usual string inspired models, it has been believed that moduli can not be thermalized due to their possible Planck-suppressed couplings with the visible sector matter and radiation fields. Unlike this, in large volume scenarios these moduli-visible sector couplings are less (string $M_s\sim\frac{M_p}{\sqrt{\cal V}}$, where ${\cal V}$ is the volume of the Calabi Yau) suppressed and consequently it has been shown \cite{CicoliFTLVS} that in LVS models some of the moduli can thermalize via the possible visible sector interaction channels sufficiently below the Planck scale. This also implies the need of investigations (in a range of temperature variation) of de-Sitter minima of zero temperature and newly appearing ones (if any) after the inclusion of finite temperature effects along with the runaway behavior of the effective potential. In this article, after investigating various modular dynamics of closed string moduli (the divisor volumes $\tau_B$ and $\tau_S$ of the Swiss-Cheese Calabi Yau we have been using), we will be interested in exploring the finite temperature effects along with addressing the issue of cosmological moduli problem.

The paper is organized as follows. In section {\bf 2}, we start with a brief summary of our D3/D7 Swiss-Cheese setup. The section {\bf 3} has a detailed calculation of moduli dynamics and couplings in which we compute the moduli masses and their various possible couplings with visible MSSM-like matter particles forming a thermal bath. Section {\bf4} is about the possible decay channels of closed string moduli and various branching ratios along with the estimates of closed string moduli decay rates in matter particles. In section {\bf5}, we discuss the finite temperature effects (up to 2-loops) to the non-supersymmetric metastable de-Sitter vacuum realized in our D3/D7 Swiss-Cheese setup in large volume limit elaborating on the possibility of moduli thermalization with the observation that de-Sitter is stable against any such finite temperature effects. Finally we provide an appendix of some intermediate steps of calculations.

\section{``Big" Divisor $D3/D7$ LVS Swiss-Cheese Setup}

In this section, we first briefly describe our D3/D7 Swiss-Cheese setup\footnote{Two nice reviews of cosmo/pheno implications of our Swiss-Cheese setup(s) have been given in \cite{MLVSReview}.} \cite{D3_D7_Misra_Shukla}: For exploring on cosmological implications of Type IIB compactifications in large volume scenarios, we started with a setup of type IIB compactified on the orientifold of a ``Swiss-Cheese Calabi-Yau" in the large volume limit. We included perturbative $\alpha^\prime$-corrections as well as its modular completion, and the instanton-generated non-perturbative superpotential which is written out respecting the (subgroup, under orientifolding, of) $SL(2,{\bf Z})$ symmetry of the underlying parent type IIB theory. With this setup, in  \cite{dSetal,largefNL_r_axionicswisscheese}, we  addressed some cosmological issues like $dS$ realization (without need of any uplifting mechanism), embedding inflationary scenarios and realizing non-trivial non-Gaussianities in the context of type IIB Swiss-Cheese Calabi Yau orientifold in LVS. This was realized with the inclusion of (non-)perturbative $\alpha^{\prime}$-corrections to the K\"{a}hler potential and non-perturbative instanton contribution to the superpotential, and without inclusion of any probe brane in the setup. Followed by this, for studying phenomenological issues in order to support MSSM (-like) models and for resolving the tension between LVS cosmology and LVS phenemenology within a string theoretic setup, a mobile space-time filling $D3-$brane and stacks of $D7$-branes wrapping the ``big" divisor $\Sigma_B$ along with magnetic fluxes is included. In this $D3/D7$ Swiss-Cheese setup, we discussed several issues of string phenomenology, like- resolution of the tension between large volume phenomenology and cosmology, realizing $ g_{YM}\sim O(1)$, and $TeV$-gravitino mass alongwith the evaluation of various soft supersymmetry breaking parameters consistent with several checks  \cite{D3_D7_Misra_Shukla,Sparticles_Misra_Shukla,MSfermionmasses}. In order to have $ g_{YM}\sim O(1)$ for supporting (MS)SM, in the previously studied LVS models with fluxed $D7$-branes, it was argued that $D7$-brane had to wrap the ``small" divisor $\Sigma_S$. Unlike this, in our setup we show the same to be possible with ``big" divisor $\Sigma_B$ due to the possible competing contributions coming from the Wilson line moduli.

Now let us review the required details of our setup, which we have just discussed above: It is type IIB compactification on the orientifold of a ``Swiss-Cheese Calabi-Yau" in the LVS limit. The Swiss-Cheese Calabi Yau we are using, is a projective variety in ${\bf WCP}^4[1,1,1,6,9]$ given as, \begin{equation}
x_1^{18} + x_2^{18} + x_3^{18} + x_4^3 + x_5^2 - 18\psi \prod_{i=1}^5x_i - 3\phi x_1^6x_2^6x_3^6 = 0
\end{equation} which has two (big and small) divisors $\Sigma_B(x_5=0)$ and $\Sigma_S(x_4=0)$.

Further working in the $x_2=1$-coordinate patch, for definiteness, we define inhomogeneous coordinates $z_1={x_1}/{x_2},\ z_2={x_3}/{x_2},\ z_3=
{x_4}/{x_2^6}$ and $z_4={x_5}/{x_2^9}$ three of which get identified with the mobile $D3$-brane position moduli. With the inclusion of (non-) perturbative $\alpha^{\prime}$ corrections and its modular completion, the K\"{a}hler potential (in appropriate $N=1$ coordinates) is given as \cite{D3_D7_Misra_Shukla}:
\begin{eqnarray}
\label{eq:K}
& & {\cal K} \biggl(\bigl{\{}T_B,\bar{T}_B;T_S,\bar{T}_S;{\cal G}^a,{\bar{\cal G}}^a;\tau,{\bar\tau}\bigr{\}};\{z_{i},{\bar z}_{i};{\cal A}_1,{\bar{\cal A}_1}\}\biggr)=
- ln[-i(\tau-{\bar\tau})] - ln[i\int_{CY_3}\Omega\wedge{\bar\Omega}]\nonumber\\
& & -2 ln\biggl[a(T_B + {\bar T}_B- \gamma K_{\rm geom}|_{\Sigma_B})^{\frac{3}{2}} -a(T_S + {\bar T}_S - \gamma K_{\rm geom}|_{\Sigma_S})^{\frac{3}{2}} + \sum_{m,n\in{\bf Z}^2/(0,0)}
\frac{({\bar\tau}-\tau)^{\frac{3}{2}}}{(2i)^{\frac{3}{2}}|m+n\tau|^3} \frac{\chi}{2 }\nonumber\\
& &   - {4\sum_{m,n\in{\bf Z}^2/(0,0)}
\frac{({\bar\tau}-\tau)^{\frac{3}{2}}}{(2i)^{\frac{3}{2}}|m+n\tau|^3}\sum_{\beta\in H_2^-(CY_3,{\bf Z})} {n^0_\beta}} Cos(mk.{\cal B} + nk.c) \biggr]
\end{eqnarray} where $n^0_\beta$ are the genus-0 Gopakumar-Vafa invariants for the holomorphic curve $\beta\in H_2^-(CY_3,{\bf Z})$, $k_a=\int_\beta\omega_a$ are the degrees of the rational curve, $\chi$ is the Euler-characteristic of the Calabi Yau, the real RR two-form potential $C_2=c_a\omega^a$; $\omega^a\in h^{1,1}_-(CY_3)$ and $z_i,{\cal A}_I$'s are set of open string moduli; the D3-brane position moduli and complexified Wilson line modulus respectively, while $K_{\rm geom}|_{B,S}$ are the geometric K\"{a}hler potentials for divisors $\Sigma_B$ and $\Sigma_S$ respectively\footnote{It has been shown in \cite{D3_D7_Misra_Shukla} that $K_{\rm geom}|_{\Sigma_B}\sim{\frac{{\cal V}^{2/3}}{\sqrt{ln{\cal V}}}}$ and $K_{\rm geom}|_{\Sigma_S}\sim{{\cal V}^{1/3}\sqrt{ln{\cal V}}}$ and hence after getting multiplied by a factor $\gamma\sim\frac{1}{{\cal V}}$, these geometric contributions to K\"{a}hler potential are subdominant in large volume limit.} and are shown to be subdominant in LVS limit \cite{D3_D7_Misra_Shukla}.
The appropriate $N=1$ coordinates \{$T_{\{B,S\}}, {\cal G}^a, S\}$ in the presence of a single mobile spacetime filling $D3$-brane and stacks of $D7$-branes wrapping the big divisor $\Sigma_B$ along with internal fluxes are defined as \cite{jockersetal},

\begin{eqnarray}
\label{eq:N=1_coords}
& & S = \tau + \kappa_4^2\mu_7{\cal L}_{A{\bar B}}\zeta^A{\bar\zeta}^{\bar B}, \ \tau=l+ie^{-\phi}; \ {\cal G}^a = c^a - \tau {\cal B}^a \nonumber\\
& & T_\alpha=\frac{3i}{2}(\rho_\alpha - \frac{1}{2}\kappa_{\alpha bc}c^b{\cal B}^c) + \frac{3}{4}\kappa_\alpha + \frac{3i}{4(\tau - {\bar\tau})}\kappa_{\alpha bc}{\cal G}^b({\cal G}^c
- {\bar{\cal G}}^c) \nonumber\\
& & + 3i\kappa_4^2\mu_7l^2C_\alpha^{I{\bar J}}a_I{\bar a_{\bar J}} + \frac{3i}{4}\delta^B_\alpha\tau Q_{\tilde{f}} + \frac{3i}{2}\mu_3l^2(\omega_\alpha)_{i{\bar j}} \Phi^i\left({\bar\Phi}^{\bar j}-\frac{i}{2}{\bar z}^{\tilde{a}}({\bar{\cal P}}_{\tilde{a}})^{\bar j}_l\Phi^l\right)
\end{eqnarray}

where
\begin{itemize}

\item
for future reference in the remainder of the paper, one defines: ${\cal T}_\alpha\equiv\frac{3i}{2}(\rho_\alpha - \frac{1}{2}\kappa_{\alpha bc}c^b{\cal B}^c) + \frac{3}{4}\kappa_\alpha + \frac{3i}{4(\tau - {\bar\tau})}\kappa_{\alpha bc}{\cal G}^b({\cal G}^c - {\bar {\cal G}}^c)$, which is appropriate coordinate in the absence of $D3/D7$ and internal fluxes implying ${\cal B}^a = b^a$.

\item
\begin{equation}
{\cal L}_{A{\bar B}}=\frac{\int_{\Sigma^B}\tilde{s}_A\wedge\tilde{s}_{\bar B}}{\int_{CY_3}\Omega\wedge{\bar\Omega}},
  \end{equation}
$\tilde{s}_A$ forming a basis for $H^{(2,0)}_{{\bar\partial},-}(\Sigma^B)$,

\item
the fluctuations of $D7$-brane in the $CY_3$ normal to $\Sigma^B$ are denoted by $\zeta\in H^0(\Sigma^B,N\Sigma^B)$, i.e., they are the space of global sections of the normal bundle $N\Sigma^B$,

\item
${\cal B}\equiv b^a - lf^a$, where the real NS-NS two-form potential
$B_2=B_a\omega^a$; $\omega^a\in h^{1,1}_-(CY_3)$ and $f^a$ are the components of elements of two-form fluxes valued in $i^*\left(H^2_-(CY_3)\right)$, the immersion map being defined as:
$i:\Sigma^B\hookrightarrow CY_3$,

\item
$C^{I{\bar J}}_\alpha=\int_{\Sigma^B}i^*\omega_\alpha\wedge A^I\wedge A^{\bar J}$, $\omega_\alpha\in H^{(1,1)}_{{\bar\partial},+}(CY_3)$ and $A^I$ forming a basis for $H^{(0,1)}_{{\bar\partial},-}(\Sigma^B)$,

\item
$a_I$ is defined via a Kaluza-Klein reduction of the $U(1)$ gauge field (one-form) $A(x,y)=A_\mu(x)dx^\mu P_-(y)+a_I(x)A^I(y)+{\bar a}_{\bar J}(x){\bar A}^{\bar J}(y)$, where $P_-(y)=1$ if $y\in\Sigma^B$ and -1 if $y\in\sigma(\Sigma^B)$,

\item
$z^{\tilde{a}}$ are $D=4$ complex scalar fields arising due to complex structure deformations of the Calabi-Yau orientifold defined via: $\delta g_{{\bar i}{\bar j}}(z^{\tilde{a}})=-\frac{i}{||\Omega||^2}z^{\tilde{a}}\left(\chi_{\tilde{a}}\right)_{{\bar i}jk}\left({\bar\Omega}\right)^{jkl}g_{l{\bar j}}$, where $\left(\chi_{\tilde{a}}\right)_{{\bar i}jk}$ are components of elements of $H^{(2,1)}_{{\bar\partial},+}(CY_3)$,

\item
$\left({\cal P}_{\tilde{a}}\right)^i_{\bar j}\equiv\frac{1}{||\Omega||^2}{\bar\Omega}^{ikl}\left(\chi_{\tilde{a}}\right)_{kl{\bar j}}$, i.e.,
${\cal P}:TCY_3^{(1,0)}\longrightarrow TCY_3^{(0,1)}$ via the transformation:
$\Phi\stackrel{\rm c.s.\ deform}{\longrightarrow}\Phi^i+\frac{i}{2}z^{\tilde{a}}\left({\cal P}_{\tilde{a}}\right)^i_{\bar j}{\bar\Phi}^{\bar j}$,

\item
$\Phi^i$ are scalar fields corresponding to geometric fluctuations of $D3$-brane inside the Calabi-Yau and defined via: $\Phi(x)=\Phi^i(x)\partial_i + {\bar\Phi}^{\bar i}({\bar x}){\bar\partial}_{\bar i}$, and

\item
$Q_{\tilde{f}}\equiv l^2\int_{\Sigma^B}\tilde{f}\wedge\tilde{f}$, where $\tilde{f}\in\tilde{H}^2_-(\Sigma^B)\equiv{\rm coker}\left(H^2_-(CY_3)\stackrel{i^*}{\rightarrow}H^2_-(\Sigma^B)\right)$.
\end{itemize}

With appropriate choice of fluxes, as in KKLT scenarios as well as in race-track scenarios with large divisor volumes (See \cite{racetrack_small_Wcs} for the latter), it is possible to tune the complex-structure moduli-dependent superpotential to be negligible as compared to the non-perturbative superpotential and hence the
superpotential in the presence of an $ED3-$instanton is of the type (See \cite{Grimm,Maldaetal_Wnp_pref,Ganor1_2}):
\begin{eqnarray}
\label{eq:W}
& & {\cal W}\sim \left(1 + z_1^{18} + z_2^{18} + z_3^2 - 3\phi_0z_1^6z_2^6\right)^{n_s}e^{in^sT_s}\Theta_{n^s}({\cal G}^a,\tau)
\end{eqnarray}
where $n^s$ is the $D3$-instanton number, the holomorphic pre-factor $\left(1 + z_1^{18} + z_2^{18} + z_3^2 - 3\phi_0z_1^6z_2^6\right)^{n_s}$ represents a one-loop determinant of fluctuations around the $ED3$-instanton and the holomorphic Jacobi theta function of index $n^s$,  $\Theta_{n^s}(\tau,{\cal G}^a)$ encodes the contribution of $D1$-instantons in an $SL(2,{\bf Z})$-covariant form, and is defined via $\Theta_{n^s}(\tau,{\cal G}^a)=\sum_{m_a}e^{\frac{i\tau m^2}{2}}e^{in^s {\cal G}^am_a}$ where ${\cal G}^a=c^a-\tau{\cal B}^a, {\cal B}^a\equiv b^a - lf^a$,  $f^a$ being the components of  two-form fluxes valued in $i^*\left(H^2_-(CY_3)\right)$.
Also a geometric realization of term $e^{in^sT_s}=e^{-n^s{\rm vol}(\Sigma_S)+i...}$ being a section of the inverse divisor bundle $n^s[-\Sigma_S]$ implies that the holomorphic prefactor $(1 + z_1^{18} + z_2^{18} + z_3^2 - 3\phi_0z_1^6z_2^6)^{n_s}$ has to be a section of $n^s[\Sigma_S]$ to compensate and the holomorphic prefactor, a section of $n^s[\Sigma_S]$ having no poles, must have zeros of order $n^s$ on a manifold homotopic to $\Sigma_S$  (See \cite{Ganor1_2}). 

The $N=1$ scalar potential at zero temperature is given as:
\be
\label{eq:VN=1}
V_{T=0} = e^{{\cal K}}\left({\cal K}^{A\bar B}{{\cal D}_A {\cal W}}{\bar{\cal D}_{\bar B} {\bar{\cal W}}}-3|{\cal W}|^2\right)
\ee

From the above equation, the most dominant contribution to the $N=1$ scalar potential in our D3/D7 Swiss-Cheese setup comes out to be a positive semi-definite term coming from the small divisor contribution $K^{\rho^s {\bar{\rho^s}}} ({\cal D}_{\rho_s}W) ({\bar{\cal D}}_{\bar{\rho_s}}{\bar W}) e^K$ in large volume limit and the same utilizing the LVS limit is given as:
\be
\label{eq:scalar_potential}
V_{T=0} = K^{\rho^s {\bar{\rho^s}}} ({\cal D}_{\rho_s}W) ({\bar{\cal D}}_{\bar{\rho_s}}{\bar W}) e^K \sim {K^{\rho^s {\bar{\rho^s}}} {\frac{{\cal V}^{\frac{n^s}{2}} e^{-2n^s\tau_s}}{a(2\tau_B + C_B)^{3/2}-a(2\tau_S + C_S)^{3/2} + C_{\alpha^{\prime}}+ C_{{n^0_{\beta}}}}}},
\ee
where
\bea
& & C_B = 3i\kappa_4^2\mu_7l^2C_B^{I{\bar J}}a_I{\bar a_{\bar J}} + \frac{3i}{4}\tau Q_{\tilde{f}} + \frac{3i}{2}\mu_3l^2(\omega_B)_{i{\bar j}} \Phi^i\left({\bar\Phi}^{\bar j}-\frac{i}{2}{\bar z}^{\tilde{a}}({\bar{\cal P}}_{\tilde{a}})^{\bar j}_l\Phi^l\right)\nonumber\\
  & & C_S = \frac{3i}{2}\mu_3l^2(\omega_S)_{i{\bar j}} \Phi^i\left({\bar\Phi}^{\bar j}-\frac{i}{2}{\bar z}^{\tilde{a}}({\bar{\cal P}}_{\tilde{a}})^{\bar j}_l\Phi^l\right)\nonumber\\
    & & C_{\alpha^{\prime}} = \frac{\chi}{2}\sum_{m,n\in{\bf Z}^2/(0,0)}
\frac{(\tau - {\bar\tau})^{\frac{3}{2}}}{(2i)^{\frac{3}{2}}|m+n\tau|^3}\nonumber\\
& & C_{n^0_{\beta}} = - 4\sum_{\beta\in H_2^-(CY_3,{\bf Z})}n^0_\beta\sum_{m,n\in{\bf Z}^2/(0,0)} \frac{(\tau - {\bar\tau})^{\frac{3}{2}}}{(2i)^{\frac{3}{2}}|m+n\tau|^3}cos\left((n+m\tau)k_a\frac{({\cal G}^a-{\bar{\cal  G}}^a)}{\tau - {\bar\tau}} - mk_a {\cal G}^a\right)\nonumber\\
\eea
where $\rho_s$ is the complexified ``small" divisor volume along with a component of RR four-form. The aforementioned scalar potential is a generalization of de-Sitter vacuum realized in our LVS cosmology setup \cite{dSetal} after the inclusion of D3/D7 brane in the setup in order to support string phenomenology in the same string theoretic setup. This is the zero-temperature effective potential and we will investigate the finite temperature effects to the same in one of the sections of this article.

Here we would like to emphasize that we have considered the rigid limit of wrapping of the $D7$-brane around $\Sigma_B$ (to ensure that there is no obstruction to a chiral matter spectrum) which is effected by considering zero sections of normal divisor bundle $N\Sigma_B$ and hence there is no superpotential generated due to the fluxes on the world volume of the $D7$-brane \cite{jockersetal}. Also the non-perturbative superpotential due to gaugino condensation on a stack of $N$ $D7$-branes wrapping $\Sigma_B$ will be proportional to $\left(1+z_1^{18}+z_2^{18}+z_3^3-3\phi_0z_1^6z_2^6\right)^{\frac{1}{N}}$ (See \cite{Maldaetal_Wnp_pref,Ganor1_2}) , which according to \cite{Ganor1_2}, vanishes as the mobile $D3$-brane touches the wrapped $D7$-brane. To effect this simplification, we will henceforth be restricting the mobile $D3$-brane to the ``big" divisor $\Sigma_B$. It is for this reason that we are justified in considering a single  wrapped $D7$-brane, which anyway can not effect gaugino condensation. We also want to mention here that even though the hierarchy in the divisor volumes (namely vol$(\Sigma_S)\sim ln {\cal V}$ and vol$(\Sigma_B)\sim {\cal V}^{\frac{2}{3}}$) was obtained in \cite{Balaetal2} in a setup without a mobile space-time filling $D3$-brane and $D7$-branes wrapping a divisor and assuming that the superpotential receives an ${\cal O}(1)$ dominant contribution from the complex structure superpotential, one can show that using the correct choice of K\"{a}hler coordinates $T_S$ and $T_B$, the divisor volumes continue to stabilize at the same values in our setup wherein the superpotential receives the dominant contribution from the non-perturbative instanton-generated superpotential \cite{Sparticles_Misra_Shukla}. Also we would like to mention one of the most crucial thing in our setup that we are assuming throughout that for appropriate choice of holomorphic isometric involution $\sigma$, the genus-zero Gopakumar-Vafa invariants $n^0_{\beta}$ are very large \cite{Curio+Spillner,Klemm_GV} and this has been very important for resolving the $\eta$- problem while studying axionic slow-roll inflation (see the second reference of \cite{largefNL_r_axionicswisscheese}).

\section{Moduli Masses and Their Various Couplings}

\subsection{Canonical Normalization and Moduli Masses}

Let us consider the K\"{a}hler moduli fluctuations around their respective stabilized VEVs corresponding to zero temperature scalar potential $V_{T=0}$ as: $\tau_{i} = \left<\tau_{i}\right> + \delta {\tau_{i}}; \ {i} \in \{B,S\}$ using which the Lagrangian can be written out in terms of fluctuations as follows:
\be
{\cal L}={{\cal K}_{i \bar j}}{\partial_{\mu}(\delta {\tau_{i}})}{\partial^{\mu}(\delta {\tau_{\bar j}})} - \left<V_{T=0}\right>- \frac{1}{2}({V_{T=0}})_{i \bar j} (\delta {\tau_i})(\delta {\tau_{\bar j}}) - {\cal O}(\delta {\tau^3}) 
\ee
The above Lagrangian can be written in terms of canonical normalized fields ${{\chi_S}}$ and ${{\chi_B}}$ along with the following constraints:
\begin{eqnarray}
\label{eq:canonical_vectors1}
& & \left(\begin{array}{c}
\delta{\tau_B} \\
\delta{\tau_S} \\
\end{array}
\right)=\left(\begin{array}{c}
v_{{\chi_S}}\\
\end{array}
\right) \frac{\chi_S}{\sqrt 2}+\left(\begin{array}{c}
v_{{\chi_B}} \\
\end{array}
\right)\frac{\chi_B}{\sqrt 2}\nonumber\\
& & {\cal K}_{i{\bar j}}(v_{\alpha})^i(v_{\beta})^j=\delta_{\alpha\beta}\ {\rm and} \  \frac{1}{2}({V_{T=0}})_{i{\bar j}}(v_{\alpha})^i(v_{\beta})^j = M_{\alpha}^2 \delta_{\alpha\beta};\  \ {\alpha,\beta} \in \{\chi_B,\chi_S\}
\end{eqnarray}
where $M_{\alpha}^2$'s and $v_{\alpha}$'s are eigenvalues and eigenvectors of mass squared matrix respectively (which can be computed from $(M^2)_{ij}=\frac{1}{2}({\cal K}^{-1})_{i{\bar k}}(V_{T=0})_{{\bar k}j}$)  along with eigenvectors $v_{\alpha}$'s normalized with K\"{a}her metric as ${\cal K}_{i{\bar j}}(v_{\alpha})^i(v_{\beta})^j=\delta_{\alpha\beta}$. Using equations (\ref{eq:K}, \ref{eq:W}, \ref{eq:VN=1}), we have the following expression for eigenvalues and (K\"{a}hler metric) normalized eigenvectors of mass squared metric:
\begin{eqnarray}
\label{eq:canonical_vectors2}
& & M_{\chi_B}^2\sim \frac{(n^s)^2 \sqrt{ln{\cal V}}}{{\cal V}^{n^s+1}}\sim \Bigl((n^s)^2 {\cal V}\sqrt{ln{\cal V}}\Bigr) m^2_{\frac{3}{2}}\nonumber\\
& & M_{\chi_S}^2\sim \frac{(n^s)^4 {ln{\cal V}}}{{\cal V}^{n^s}}\sim \Bigl({\cal V}^2 (n^s)^4 {ln{\cal V}}\Bigr) m^2_{\frac{3}{2}}\nonumber\\
& & \left(\begin{array}{c}
\delta{\tau_B} \\
\delta{\tau_S} \\
\end{array}
\right)\sim\left(\begin{array}{c}
\frac{{{\cal V}^{\frac{17}{36}}}}{{(ln{\cal V})}^{\frac{1}{4}}}\\
{{\cal V}^{\frac{1}{2}}}{(ln{\cal V})}^{\frac{1}{4}}\\
\end{array}
\right) \frac{\chi_S}{\sqrt{2}}+\left(\begin{array}{c}
 -{\cal V}^{\frac{37}{72}} \\
{{\cal V}^{-\frac{11}{24}}}\\
\end{array}
\right)\frac{\chi_B}{\sqrt {2}}\nonumber\\
\end{eqnarray}

Here it is interesting to see that although flcuations with respect to small divisor modulous $\delta \tau_S$ is approximately one of the canonically normalized field $\chi_S$, however there is a significant mixing between the moduli fluctuations $\delta \tau_B$ and $\delta \tau_S$ for the other canonically normalized field $\chi_B$ which plays crucial role while studying the finite temperature corrections to realized zero-temperature local de-Sitter minimum. Also we find that both divisors are heavier than the gravitino mass and for Calabi Yau Volume ${\cal V}\sim 10^6 l_s^6$ and D3-instanton number $n^s=2$ \footnote{The cosmological/astrophysical observations prefer Calabi Yau Volume ${\cal V}\sim 10^6 l_s^6$ and also for realizing de-Sitter solutions in our LVS Swiss-Cheese cosmology setup as well as in various cosmo/pheno implications, we have been using D3-instanton number $n^s=2$ along with Calabi Yau Volume ${\cal V}\sim 10^6 l_s^6$ \cite{dSetal,D3_D7_Misra_Shukla,MSfermionmasses}.}, these masses are $M_{\chi_S}\sim 10^9 {\rm TeV}$ and $M_{\chi_B}\sim 10^6 {\rm TeV}$, which are heavy enough\footnote{Unlike \cite{AstroLVS}, in which ``big" divisor is very light ($\sim MeV$) with other one being heavy, in our setup we find that both are quite heavy which is due to the possible competing contribution coming from Wilson line moduli. These large masses also imply very small life times and hence no cosmological problem.} to avoid any ``cosmological moduli problem" as they decay very fast in early times and would not spoil structure formations in post-inflationary era.

\subsection{Couplings to Gauge Bosons}
In this subsection, assuming that some (MS)SM-(like) model could be supported on the world volume of D7-brane wrapping the big divisor $\Sigma_B$, we discuss the coupling of these closed string moduli with the observable sector (MS)SM matter fields. The information of coupling of gauge Bosons to closed string moduli is encoded in gauge kinetic function $f_\alpha$ via a term ${\cal L}_{\rm gauge}$ in the Lagrangian defined as:
\be
\label{eq:Lgauge}
{\cal L}_{\rm gauge}\sim\frac{1}{g_{YM}^2}F_{\mu \nu}F^{\mu \nu}\sim Re(T_\alpha)F_{\mu \nu}F^{\mu \nu}; \ M_p = 1 \ {\rm units}
\ee
As in our $D3/D7$ Swiss-Cheese orientifold setup, (MS)SM is supported on the world volume of D7-brane wrapping the big divisor $\Sigma_B$, the gauge couplings $g_a$'s are generically defined through gauge kinetic functions $f_a$'s as:
\be
f_a = \frac{T_{\rm (MS)SM}}{4\pi} + h_a(F)S;\Longrightarrow \frac{1}{g_{a}^2}\sim Re({T_B})\sim {\cal O}(1)
\ee
Here order one gauge couplings are realized in dilute flux approximation (which are well compatible to be assumed in the context of large volume scenarios) and the difference in gauge couplings of different gauge groups in (MS)SM could be realized by turning on the internal fluxes on two cycles in the world volume of D7-brane wrapping the big divisor $\Sigma_B$. It has been observed in \cite{D3_D7_Misra_Shukla} that by constructing appropriate odd harmonic one-form, Wilson line moduli contribution to the shift in $N=1$ coordinate $T_B$ is competing with the big divisor volume $\tau_B$ and equation (\ref{eq:Lgauge}) reduces to
\be
\label{eq:bosoncoupls}
\hskip -1cm {\cal L}_{\rm gauge}\sim \frac{1}{M_p}\Biggl[\tau_B +Re\Bigl[{3i\kappa_4^2\mu_7l^2C_\alpha^{I{\bar J}}a_I{\bar a_{\bar J}} + \frac{3i}{4}\delta^B_\alpha\tau Q_{\tilde{f}} + \frac{3i}{2}\mu_3l^2(\omega_\alpha)_{i{\bar j}} \Phi^i\left({\bar\Phi}^{\bar j}-\frac{i}{2}{\bar z}^{\tilde{a}}({\bar{\cal P}}_{\tilde{a}})^{\bar j}_l\Phi^l\right)}\Bigr]\Biggr]F_{\mu \nu}F^{\mu \nu}\nonumber\\
\ee
Now considering the fluctuations $\tau_B = \left<\tau_B\right>+\delta\tau_B$ and using the result \cite{D3_D7_Misra_Shukla} that there is a possible cancelation of big divisor volume with Wilson line moduli contribution and D3-brane position moduli to be stabilized around $\Phi_i\equiv z_i\sim {\cal V}^{\frac{1}{36}}$ along with canonical normalized eigenvectors from (\ref{eq:canonical_vectors2}), we have the following interaction lagrangian in dilute flux approximations:
\bea
& & {\cal L}_{{\chi_B} \gamma \gamma}\sim {\frac{{\cal V}^{\frac{33}{72}}}{M_p}} {\chi_B G_{\mu \nu} G^{\mu \nu}}\nonumber\\
& & {\cal L}_{{\chi_S} \gamma \gamma}\sim {\frac{{\cal V}^{\frac{5}{12}} {(ln{\cal V})}^{-\frac{1}{4}}}{M_p}} {{\chi_S} G_{\mu \nu} G^{\mu \nu}}
\eea
where $G_{\mu \nu}={\sqrt{\left<\tau_B\right>}}F_{\mu \nu}$. From above, we can read out the couplings of canonical fields ${\chi_B}$ and ${\chi_S}$ with gauge fields as below:
\be
\label{eq:bosoncouplings}
\lambda_{{\chi_B} \gamma \gamma}\sim {\frac{{\cal V}^{\frac{33}{72}}}{M_p}} ; \ \   \lambda_{{\chi_S} \gamma \gamma}\sim {\frac{{\cal V}^{\frac{5}{12}} {(ln{\cal V})}^{-\frac{1}{4}}}{M_p}}.
\ee
Thus it is interesting to observe that coupling of muduli to gauge bosons is suppressed almost of the order of string which is much below the Planck scale in large volume limit. Also we find that although the (MS)SM-like model is supported on stacks of D7-brane wrapping the ``big" divisor $\Sigma_B$, however there is a significant (non-zero) coupling of canonically normalized field $\chi_S$ (which is mostly aligned along $\tau_S$ direction) with the photon. Further these large coupling of divisor volume moduli with photons play crucial role for thermalization processes which we will discuss in one of the coming sections. This is interesting to recall here that large couplings of divisor volume moduli with observable sector fields are one of the novel features of large volume scenarios (as also observed in \cite{AstroLVS}) which in unlike the other string inspired models like KKLT.

\subsection{Couplings to Matter Fermions/Higgs}
In an recent work \cite{MSfermionmasses}, we have shown the possibility of realizing first two generation fermion masses as well as their neutrino masses in our setup in which the Wilson line moduli $a_I$'s are identified with the squark/slepton and the masses of their corresponding fermionic superpartners ${\cal \chi}_I$'s are the first two generation fermion masses. In this subsection, we are going to compute the coupling of matter fermions\footnote{We will be using the notations of \cite{MSfermionmasses}, ${\cal \chi}_I$ - the fermionic superpartner of Wilson line moduli $a_I$; ${\cal A}_I =$ Wilson line moduli superfields $=a_I+\theta \chi_I+...$.} ${\cal \chi_I}$ with closed string moduli which can be read off from the the fluctuations in the following Lagrangian:
\be
{\cal L}_{\rm ferm-Higgs}={{K_{{\cal \chi}_I {\bar{{\cal \chi}_J}}}{\overline{{\cal \chi}_I}}i {\gamma^{\mu}}{\partial_{\mu}{\cal \chi}_J}}+ {K_{{\cal Z} {\bar{{\cal Z}}}}{\partial_{\mu}{\cal Z}}}{\partial^{\mu}{\bar {\cal Z}}}}+ {e^{\hat{K}/2} Y_{\alpha IJ} {\cal Z}^{\alpha} \overline{{\cal \chi}_I}{\cal \chi}_J}+{{\cal L}_{\rm {Higgs-int}}}
\ee
where  the fermion bilinear terms  as well as Higgs interaction terms in the 4D effective action are generated through $\int d^4 x \, e^{\hat{K}/2} (\partial_{I}\partial_{J} {\cal W}) {\cal \chi}_I{\cal \chi}_J$as it has been argued in \cite{Sparticles_Misra_Shukla} that the Wilson line moduli ${\cal A}_I$'s could be identified with squarks/sleptons while D3-brane position moduli could be identified with two Higgses in our D3/D7 Swiss-Cheese orientifold setup, we have\footnote{The details of double derivatives of K\"{a}hler potential are given in Appendix A.}
\bea
& & {{K}}_{{\cal \chi}_I {\bar {\cal \chi}_J}}\equiv{\cal K}_{{\cal A}_I {\bar {\cal A}_J}} \sim {{\cal V}^{\frac{65}{36}}}{\cal K}_0\Bigl(1-\frac{\delta\tau_B}{2\mu_3{\cal V}^{\frac{1}{18}}}\Bigr); \ {K_{{\cal Z} {\bar{{\cal Z}}}}}\sim {{\cal V}^{\frac{1}{36}}}{\cal K}_0\bigl[1+{\cal O}(1)\delta\tau_S-{\cal O}(1)\delta\tau_B\bigr]\nonumber\\
& & e^\frac{{\hat{K}}}{2} \sim {\cal K}_0\bigl[1+{\cal O}(1){\cal K}_0\delta\tau_S-{\cal O}(1){\cal K}_0\delta\tau_B\bigr] \ {\rm where} \  {\cal K}_0=\frac{1}{{\sum_{\beta\in H^2_{{\bar\partial},-}(CY_3)}n^0_{\beta}}}\nonumber\\
\eea
After considering the canonical normalized fields and generating fermion masses through Higgs(like) mechanism and using Higgs fluctuations ${\cal Z}_i=\left<{\cal Z}_i\right>+\delta{\cal Z}_i$, we have
\bea
\label{eq:Lhiggs}
& & \hskip -1cm {\cal L}={\overline{{\cal \chi}_I}}\Bigl(i \gamma^{\mu}\partial_{\mu}+m_{{\cal \chi}}\Bigr){\cal \chi}_J -{\cal O}(1) \delta\tau_B \Bigl[{\overline{{\cal \chi}_I}}(i \gamma^{\mu}\partial_{\mu}+m_{{\cal \chi}}){\cal \chi}_J\Bigr] + \Bigl[{\cal O}(1) \delta\tau_S-{\cal O}(1)\delta\tau_B\Bigr]m_{{\cal \chi}}{\overline{{\cal \chi}_I}}{\cal \chi}_J+{\cal L}_{\delta{\cal Z}_i}\nonumber\\
& & {\cal L}_{\delta{\cal Z}_i}= {\hat{Y}_{\alpha I J}}\Bigl[1-{\cal O}(1){\cal K}_0\delta\tau_S+{\cal O}(1){\cal K}_0\delta\tau_B\Bigr]{\delta{\cal Z}_i}{{\overline{\cal \chi}_I}{\cal \chi}_J}
\eea
where without notational changes ${\cal \chi}_I$'s are canonical normalized matter fermions now and the contribution in the second term vanishes for on-shell condition while the possible fermion masses realized, are  given as eigenvalues of fermion mass matrix\footnote{In \cite{MSfermionmasses}, we have shown the possibility of realizing all first two generation fermion masses in our D3/D7 Swiss-Cheese setup for calabi yau volume ${\cal V}\sim (5\times10^5 - 10^6)$.} $m_{{\cal \chi}}= {\hat{Y}_{\alpha \beta \gamma}}\left<{\cal Z}_i\right>$ \cite{MSfermionmasses} which implies that the moduli-fermion and moduli-fermion-Higgs couplings are as follows:
\bea
& & \delta{\cal L}_{\chi_S {\overline{\cal \chi}_I}{\cal \chi}_I}\sim \left[{\cal V}^{\frac{1}{2}} (ln{\cal V})^{\frac{1}{4}}m_{{\cal \chi}}\right] {\chi_S {\overline{\cal \chi}_I}{\cal \chi}_I}\nonumber\\
& & \delta{\cal L}_{\chi_S {\overline{\cal \chi}_I}{\cal \chi}_I {\delta{\cal Z}_i}} \sim {\hat{Y}_{\alpha IJ}} {\cal K}_0 \left[{\cal V}^{\frac{1}{2}} (ln{\cal V})^{\frac{1}{4}}m_{\psi_c}\right] {\chi_S {\overline{\cal \chi}_I}{\cal \chi}_I {\delta{\cal Z}_i}} \nonumber\\
& &\delta{\cal L}_{\chi_B {\overline{\cal \chi}_I}{\cal \chi}_I}\sim \left[{\cal V}^{\frac{37}{72}} m_{{\cal \chi}}\right] {\chi_B {\overline{\cal \chi}_I}{\cal \chi}_I}  \nonumber\\
& &  \delta{\cal L}_{\chi_B {\overline{\cal \chi}_I}{\cal \chi}_I {\delta{\cal Z}_i}}\sim {\hat{Y}_{\alpha\beta\gamma}}{\cal K}_0\left[{\cal V}^{\frac{37}{72}} m_{{\cal \chi}}\right] {\chi_B {\overline{\cal \chi}_I}{\cal \chi}_I{\delta{\cal Z}_i}}\nonumber\\
\eea

\subsection{Moduli Self Couplings}

Moduli self interaction can be estimated from the trilinear term in the expansion of $N=1$ scalar potential around stabilized VEVs of moduli:
\be
V(\delta {\tau_{B}},\delta {\tau_{S}})=\left<V({\tau_{B}},{\tau_{S}})\right> + \frac{1}{2} \biggl(\frac{\partial^2 V}{\partial{\tau_{i}}\partial{\tau_{j}}}\biggr)_{min}{\delta\tau_{i}}{\delta\tau_{j}}+ \frac{1}{3!} \biggl(\frac{\partial^3 V}{\partial{\tau_{i}}\partial{\tau_{j}}\partial{\tau_{k}}}\biggr)_{min}{\delta\tau_{i}}{\delta\tau_{j}}{\delta\tau_{k}}+{\cal O}({\tau^4})
\ee
Now using general expression for $N=1$ scalar potential and considering canonically normalized fluctuations, we have the following self coupling terms in the fluctuation Lagrangian,
\bea
\label{eq:selfcouplings}
& & \delta{{\cal L}}_{\chi_S^3}\sim \Biggl[\frac{(ln{\cal V})^{\frac{5}{4}}}{{\cal V}^{n^s-\frac{3}{2}}\ (\sum_{\beta\in H^2_{{\bar\partial},-}(CY_3)}n^0_{\beta})}\Biggr]{\chi_S^3} \nonumber\\
& &  \delta{{\cal L}}_{\chi_B^3}\sim \Biggl[-\frac{(ln{\cal V})^{\frac{1}{2}}}{{\cal V}^{n^s-\frac{109}{72}}\ (\sum_{\beta\in H^2_{{\bar\partial},-}(CY_3)}n^0_{\beta})^2}\Biggr] {\chi_B^3}\nonumber\\
& & \delta{{\cal L}}_{\chi_S^2 \chi_B}\sim \Biggl[-\frac{(ln{\cal V})^{\frac{1}{2}}}{{\cal V}^{n^s-\frac{105}{72}}\ (\sum_{\beta\in H^2_{{\bar\partial},-}(CY_3)}n^0_{\beta})^2}\Biggr]{\chi_S^2 \chi_B} \nonumber\\
& & \delta{{\cal L}}_{\chi_B^2 \chi_S}\sim \Biggl[\frac{(ln{\cal V})^{\frac{1}{2}}}{{\cal V}^{n^s-\frac{52}{36}}\ (\sum_{\beta\in H^2_{{\bar\partial},-}(CY_3)}n^0_{\beta})^2}\Biggr] {\chi_B^2 \chi_S}\nonumber\\
\eea
As we are assuming throughout the setup that for appropriate choice of holomorphic isometric involution $\sigma$, the genus-zero Gopakumar-Vafa invariants $n^0_{\beta}$ are very large \cite{Curio+Spillner,Klemm_GV} (and we have been consistently using D3-instanton number $n^s=2$ in our setup) implying that moduli self couplings are negligible in large volume limit. As the interaction rates of these closed string moduli ($\chi_B$ and $\chi_S$) are proportional to the coupling squares, these highly suppressed self-couplings estimated above closes all the channels involving their self-interactions.

\section{Moduli Decay Channels to Matter Fields}
The moduli life times depend on two factors, first being their masses and the second being the various couplings with species involved in the interaction process, and the life time is inversely proportional to some powers of masses and couplings. In this section we elaborate on the possible decay channels through which moduli would have been thermalized by interacting with some (MS)SM-like matter fields. The moduli decay rates are given as \cite{AstroLVS,CicoliFTLVS},
\bea
& & \Gamma_{\left\{\chi_B,\chi_S\right\}\longrightarrow{\gamma\gamma}}=\frac{{\lambda_\gamma}^2 M_{\left\{\chi_B,\chi_S\right\}}^3}{64\pi M_p^2}\nonumber\\
& &  \Gamma_{\left\{\chi_B,\chi_S\right\}\longrightarrow{{\overline{\cal \chi}_I}{\cal \chi}_J}}=\frac{{\lambda_{{\cal \chi}}}^2 m_{{\cal \chi}}^2 M_{\left\{\chi_B,\chi_S\right\}}}{8 \pi M_p^2}\left[1-\frac{4 m_{{\cal \chi}}^2}{M_{\left\{\chi_B,\chi_S\right\}}^2}\right]^{\frac{3}{2}}
\eea
where $\lambda_{\gamma}$ and $\lambda_{{\cal \chi}}$ are moduli couplings to photons and (canonically normlized) matter fermions ${{\cal \chi}_I}$ and, $m_{{\cal \chi}}$ and $M_{\left\{\chi_B,\chi_S\right\}}$ are masses of respective fermion and moduli which are decaying in that fermion channel. Further, as in an expanding universe for a particle species to be in thermal equilibrium with a thermal bath at temperature T, its interaction rate with the other particles in the bath must be greater than the expansion rate of the universe (encoded in Hubble parameter ${\cal H}$), i.e.
\be
\label{eq:decayH}
\Gamma_{\rm interaction} \gg {\cal H} \sim \frac{N_p^{\frac{1}{2}} T^2}{M_p};
\ee
where  $N_p$ is the total number of parameters in the therory (e.g. for MSSM $N_p\sim 100$). Now moduli can thermalize via several interaction process, for example $2\longleftrightarrow2$ processes (which involve process like scattering, annihilation, inverse pair production etc) and $1\longleftrightarrow2$ processes (which involve decay, inverse decay, single particle production etc.) and one has to explore which among these processes are compatible with (\ref{eq:decayH}) otherwise those channels are ruled out from the possibility of thermalizing moduli.

Before investigating the possible allowed interaction process for moduli thermalization in our setup, let us first briefly summarize the various channel showing possibility of moduli thermalization in usual large volume setup \cite{CicoliFTLVS}. The thermally averaged interaction rate $\left<\Gamma\right>$ of an interaction process depends on the interaction time which is estimated by number densities of species involved, effective cross-section $\sigma$ and relative velocities between the particles implying $\left<\Gamma\right>\sim \sigma T^3$ (using number density $\sim T^3$) for relativistic particles. After estimating the effective cross-sections in various classes of $2\longleftrightarrow2$ and $1\longleftrightarrow2$ processes (like process with only gauge vertices, with gravitational as well as gauge vertices, only gravitational vertices), and imposing the constraint (\ref{eq:decayH}), one can compute the various temperature bounds for searching a suitable channel of moduli thermalization and ruling out the irrelevant ones. The results of \cite{CicoliFTLVS} are summarized in the first two tables (Table[1] and Table[2]).

\begin{table}
\begin{center}
\begin{tabular}{|c|c|c|}
\hline
 & & \\
$2\longleftrightarrow2$ & Interaction rates ($\Gamma$) & Interaction rates ($\Gamma$)\\
& (Renormalizable interaction) & (Non-renormalizable interaction) \\
& & \\
\hline
\hline
 & & \\
 & $\left<\Gamma\right>\sim \alpha^2 \frac{T^5}{(T^2+M_m^2)^2}$ & (a). 1-gravity + 1-gauge vertices\\
& & $\left<\Gamma\right>\sim {\sqrt {\cal D}} g^2 \frac{T^3}{M_p^2}$ \\
& & \\
 & $M_m$=Mass of the particle mediating the  & (b). 2-gravitational vertices \\
 & process; $\alpha\sim\frac{g^2}{4\pi}$, $g$ being gauge coupling. & $\left<\Gamma\right>\sim {\cal D} \frac{T^5}{M_p^4}$ \\
  & & \\
\hline
& & \\
Temperature &$T < \alpha^2 N_p^{1/2} M_p$; \  for $M_m=0$ & $T > \frac{N_p^{1/2}M_p}{g^2 \sqrt{\cal D}}$, \ for (a) above.\\
bounds & $\Bigl(\frac{N_p^{1/2}M_m^4}{\alpha^2 M_p}\Bigr)^{1/3} < T $; \  for $M_m\neq0$ & $T > N_p^{1/6} \frac{M_p}{{\cal D}^{1/3}}$, \ for (b) above.\\
 & & \\
\hline
\end{tabular}
\end{center}
\caption{The second column in the table above corresponds to ``Renormalizable interaction" while the third one corresponds to ``Non-renormalizable interactions".}
\end{table}

\begin{table}
\begin{center}
\begin{tabular}{|c|c|c|}
\hline
 & & \\
$1\longleftrightarrow2$ & {Interaction rates ($\Gamma$)}&  Interaction rate ($\Gamma$) \\
& (of renormalizable processes & (of gravity or renormalizable \\
& with massless or masses & process for mass $M_m$ \\
& $M_m$ mediators; $T > M_m$ )& mediators with $T < M_m$)\\
& & \\
\hline
\hline
& & \\
Decay& $\left<\Gamma_D\right>\sim g^2\frac{m^2}{T}; T\geq m$ & $\left<\Gamma_D\right>\sim \sqrt{\cal D}\frac{m^4}{M_m^2 T}; T\geq m$ \\
& \hskip 0.8cm $\sim g^2{m}; T\leq m$ & \hskip 0.8cm $\sim \sqrt{\cal D}\frac{m^3}{M_m^2}; T\leq m$ \\
& & \\
Inverse Decay& $\left<\Gamma_{ID}\right>\sim g^2\frac{m^2}{T}; T\geq m$ & $\left<\Gamma_{ID}\right>\sim \sqrt{\cal D}\frac{m^4}{M_m^2 T}; T\geq m$ \\
& $\left<\Gamma_{ID}\right>\sim g^2{m}(\frac{m}{T})^{3/2} e^{-m/T}; T\leq m$ & $\left<\Gamma_{ID}\right>\sim \sqrt{\cal D}\frac{m^3}{M_m^2}(\frac{m}{T})^{3/2} e^{-m/T}; T\leq m$ \\
& m=mass of decaying particle& \\
  & & \\
\hline
& & \\
Temperature &$m < T < \Bigl(\frac{g^2 m^2 M_p}{ N_p^{1/2}}\Bigr)^{1/3}$; \  for $M_m=0$ & $T > \frac{N_p^{1/2}M_p}{g^2 \sqrt{\cal D}}$, \ for (a) above.\\
bounds & $\Bigl(\frac{N_p^{1/2}M_m^4}{\alpha^2 M_p}\Bigr)^{1/3} < T $; \  for $M_m\neq0$ & $T > N_p^{1/6} \frac{M_p}{{\cal D}^{1/3}}$, \ for (b) above.\\
 & & \\
\hline
\end{tabular}
\end{center}
\caption{In this table the second column corresponds to renormalizable interaction with massless or mass $M_m$ mediators with $T > M_m$ while the third column corresponds to gravity or renormalizable interactions for mass $M_m$ mediators with $T < M_m$.}
\end{table}

With the pieces of information in the tables ([1] and [2]) and various masses and couplings computed in our setup listed in table [3], now we start investigating the possible process(es) which could help in moduli thermalization in our D3/D7 Swiss-Cheese large volume setup. A particular process is allowed if it is consistent with (\ref{eq:decayH}). We observe that for the given divisor volume moduli masses $M_{\chi_B}\sim {{\cal V}^{-\frac{n^s+1}{2}}}\  M_p; \ M_{\chi_S}\sim {{\cal V}^{-\frac{n^s}{2}}}\ M_p$, and the gauge couplings $g_{YM}\sim{\cal O}(1)$ and moduli-photon couplings $\lambda_{{\chi_B} \gamma \gamma}\sim {\frac{{\cal V}^{\frac{33}{72}}}{M_p}} ; \ \   \lambda_{{\chi_S} \gamma \gamma}\sim {\frac{{\cal V}^{\frac{5}{12}} {(ln{\cal V})}^{-\frac{1}{4}}}{M_p}}$ implying that at each gravitational vertex appearing, there will be a dimensionless constant $\sqrt{\cal D}$ factor given asrespectively\footnote{Intuitively we can also argue that large ${\cal D}$ values are expected in large volume scenarios as: due to existence of a new large scale (the big divisor volume itself) the gravitational interactions involved, are string mass suppressed $M_s\sim\frac{M_p}{\sqrt{\cal V}}$ instead of Planck mass suppressed and hence each gravitation vertex contributes one ${\cal V}$ factor to interaction rate $\left<\Gamma\right>$ from $\frac{1}{M_s^2}$ suppression instead of $\frac{1}{M_p^2}$ implying the dimensionless geometric constant ${\cal D}\sim {\cal V}^2$. Here one should keep in mind that $\cal D$ dimensionless constant corresponds to processes involving two-gravitational vertices and, for one gauge and one gravitational vertices, it will appear as $\sqrt{\cal D}$.},
\be
\label{eq:calD}
{\cal D}_{\chi_B}\sim {{\cal V}^{\frac{11}{6}}} ; \ \   {\cal D}_{\chi_S}\sim {\frac{{\cal V}^{\frac{5}{3}} }{{(ln{\cal V})}}}.
\ee
Now assuming that the number of degrees of freedoms $N_p\sim {\cal O}(100)$ in some (MS)SM-like model supported on stacks of D7-brane wrapping the big divisor in our setup and using (\ref{eq:calD}) along with Table[1] and Table[2], we come up with simple estimates similar to the results of \cite{CicoliFTLVS} that the divisor volume moduli $\tau_B$ and $\tau_S$ can thermalize at a temperature sufficiently below the Planck scale. This is a novel consequence of large volume scenarios and does not hold for KKLT(-type) models, as in those models the geometric dimensionless constant appearing from gravitational vertex ${\cal D}\sim{\cal O}(1)$ implying the temperature upper bound to go beyond (or comparable to) Planck scale and hence more away from supergravity regime and hence spoiling the possibility of moduli thermalization via some of $2\longleftrightarrow2$ and $1\longleftrightarrow2$ processes. These moduli decay processes reheat the universe, generating large amounts of entropy and could possibly dilute any primordial baryon asymmetry. However such issues have been resolved in models with thermal inflation \cite{Thermalinflation2} and hence generating required baryon asymmetry is not that big issue. The other problematic possibility with heavy divisor volume moduli $M_{\chi_S}\sim 10^9 {\rm TeV}$ and $M_{\chi_B}\sim 10^6 {\rm TeV}$ in our setup, is the possible gravitino overproduction and reheating related issues. In order to estimate the gravitino channel branching ratio let us look at the responsible moduli-gravitino couplings.

The moduli decay to gravitinos ($\chi_B, \chi_S\longrightarrow 2\Psi$ channel) is dictated by the couplings appearing in a term $e^{K/2}|W|{\bar\Psi_\mu}[\gamma^\mu,\gamma^\nu]{\Psi_\nu}$ in the Lagrangian. Considering the fluctuations in the closed string moduli $\tau_{B,S}$\footnote{The interaction term with divisor volume fluctuations appears from K\"{ahler} potential and superpotential which is $e^{K/2}|W|\Bigl\{\partial_{\tau_S}(K + ln|W|^2)\delta{\tau_S}+\partial_{\tau_B}(K + ln|W|^2)\delta{\tau_B}\Bigr\} {\bar\Psi_\mu}[\gamma^\mu,\gamma^\nu]{\Psi_\nu} $.} and utilizing their canonical normalized forms result in an interaction term as following:
\be
\delta{\cal L}\sim e^K |W|\left\{{{\cal V}^{\frac{1}{2}}(ln{\cal V})^{\frac{1}{4}}} \chi_S +\frac{{\cal V}^{\frac{37}{72}}}{{\sum_{\beta\in H^2_{{\bar\partial},-}(CY_3)}n^0_{\beta}}} \chi_B\right\}{\bar{\Psi}}_\mu[\gamma^\mu,\gamma^\nu]{\Psi_\nu}
\ee
One can read out the moduli-gravitino couplings from the above result to be
\bea
& & \lambda_{\bar{\Psi}\Psi\chi_B}\sim \frac{{\cal V}^{\frac{37}{72}}}{{\sum_{\beta\in H^2_{{\bar\partial},-}(CY_3)}n^0_{\beta}}}\nonumber\\
& & \lambda_{\bar{\Psi}\Psi\chi_S}\sim {{\cal V}^{\frac{1}{2}}(ln{\cal V})^{\frac{1}{4}}}.
\eea
As we are assuming throughout the setup that for appropriate choice of holomorphic isometric involution $\sigma$, the genus-zero Gopakumar-Vafa invariants are very large with $n^0_{\beta}\sim\frac{{\cal V}}{10}$, we observe that the Branching ratios of $\chi_B$  decaying into gravitinos is suppressed by large volume factor for ``big" divisor $Br(\chi_{B}\rightarrow{\bar\Psi}\Psi)\sim {\sim{\frac{100}{{\cal V}^\frac{19}{9}}}}\sim 10^{-10}$ (which is similar to \cite{AstroLVS}), however for the ``small" divisor $Br(\chi_{S}\rightarrow{\bar\Psi}\Psi)\sim {\cal O}(1)$ similar to the usually expected one order one branching ratio \cite{Gravitinoprodcuction} and the same could be resolved by some proposals like \cite{Thermalinflation1}. A list of various moduli-visible sector couplings and moduli masses is given in Table[3].

\begin{table}
\begin{center}
\begin{tabular}{|c|c|c|}
\hline
 & & \\
& $\chi_B$ &  $\chi_S$ \\
 & & \\
\hline
\hline
 & & \\
Mass & $M_{\chi_B}\sim n^s{\cal V}^{-\frac{n^s+1}{2}}$& $M_{\chi_S}\sim (n^s)^2{\cal V}^{-\frac{n^s}{2}}$ \\
 & & \\
\hline
 & & \\
Matter Couplings & $\lambda_{\gamma\gamma\chi_B}\sim {\cal V}^{\frac{33}{72}}$ & $\lambda_{\gamma\gamma\chi_S}\sim {\frac{{\cal V}^{\frac{5}{12}}}{(ln{\cal V})^{{\frac{1}{4}}}}}$ \\
& $\lambda_{\bar{{\cal \chi}_I}{\cal \chi}_I\chi_B}\sim {\cal V}^{\frac{37}{72}}$ & $\lambda_{\bar{{\cal \chi}_I}{\cal \chi}_I\chi_S}\sim {\cal V}^{\frac{1}{2}}$ \\
& $\lambda_{\bar{\Psi}\Psi\chi_B}\sim \frac{{\cal V}^{\frac{37}{72}}}{{\sum_{\beta\in H^2_{{\bar\partial},-}(CY_3)}n^0_{\beta}}}$ & $\lambda_{\bar{\Psi}\Psi\chi_S}\sim {{\cal V}^{\frac{1}{2}}}$ \\
 & & \\
\hline
 & & \\
Decay Modes & & \\
$\gamma \gamma$  & $\tau \sim 10^{-17}$ s &$\tau \sim 10^{-26}$s\\
$e^{+} e^{-}$ & $\tau \sim 10^{5}$s &$\tau \sim 10^{3}$s\\
$q \bar{q}$ & $\tau \sim 10^{5}$s &$\tau \sim 10^{3}$s\\
$\Psi \Psi$ & $\tau \sim 10^{-8}$s &$\tau \sim 10^{-26}$s\\
 & & \\
\hline
\end{tabular}
\end{center}
\caption{Moduli masses, various couplings and decay rates}
\end{table}

\section{$D3/D7$ Finite Temperature Corrections In LVS}
\label{D3D7FTC}
After investigating moduli masses and their couplings to (MS)SM matter fields, so far we have elaborated on the various possible moduli thermalization processes via moduli interaction with thermal matter bath of visible sector fields. Finite temperature issues have been considered with deep attention since long (see \cite{Buchmuller1,Buchmuller2,FirstFT} and references therein). Now we focus on the study of the finite temperature corrections to $N=1$ scalar potential in large volume limits  and investigate the possibility of destabilization of the local physical minimum realized in zero temperature limit. This possibility arises due to significant couplings of closed string moduli with a thermal matter bath of observable sector fields of some (MS)SM-like models supported on D7-branes wrapping the big divisor $\Sigma_B$, which results in the $T^4$ contribution to the scalar potential and for high enough temperatures this could modify/destabilise the zero-temperature minimum. In the limit of high temperature expansion ($T\gg \{M_{\rm soft},M_{\rm moduli},M_{\rm modulino}\}$), the total $N=1$ scalar potential upto two-loop finite temperature correction is given as \cite{CicoliFTLVS,FTformulae}
\be
\label{eq:V_T}
V_{\rm Tot} = V_{T=0} + V_T^{1-loop} + V_T^{2-loop} +.....
\ee
where
\begin{eqnarray}
\label{eq:V012}
& & \hskip -0.5cm V_{T=0}\sim e^KG^{\tau^\alpha{\bar\tau}^\alpha}D_{\tau^\alpha}W{\bar D}_{{\bar\tau}^\alpha}{\bar W}\sim\frac{(n^s)^2|W|^2{\cal V}^{\frac{19}{18}}}{{\cal V}^2}\sim{\cal V}^{-n^s - 2 + \frac{19}{18}}\sim {\cal V}^{\frac{19}{18}}m^2_{\frac{3}{2}}\nonumber\\
& & \hskip -0.5cm V_T^{1-loop}=-\frac{\pi^2 T^4}{90} \left( g_B+ \frac{7}{8} g_F
\right)+\frac{T^2}{24}\left(M_{\chi_B}^2+M_{\chi_S}^2+M_{\tilde{\chi_B}}^2+M_{\tilde{\chi_S}}^2+\sum_i
M_{\rm{soft},i}^2\right)+...\nonumber\\
& & \hskip -0.5cm V_T^{2-loops}=T^4\left(\kappa_1 g_{MSSM}^2+\kappa_2 \lambda_{\chi_B \gamma\gamma}^2
M_{\chi_B}^2+\kappa_3 \lambda_{\chi_S \gamma\gamma}^2
M_{\chi_S}^2+\kappa_4 \lambda_{\tilde{\chi_B}\tilde{\gamma}\gamma}^2
M_{\tilde{\chi_B}}^2+\kappa_5 \lambda_{\tilde{\chi_S}\tilde{\gamma}\gamma}^2
M_{\tilde{\chi_S}}^2+...\right)+...\nonumber\\
\end{eqnarray}
The first term in the above effective $N=1$ scalar potential (\ref{eq:V_T}) is the zero-temperature contribution obtained after considering modular completed expression (which includes (non-) perturbative $\alpha^{\prime}$ corrections) for K\"{a}hler potential and instanton-generated contribution to superpotential written out utilizing $SL(2,Z)$ (or a subgroup of the same after orientifolding) symmetry of underlying type IIB. The second term is one-loop finite temperature corrections which results by treating the species in thermal bath as non-interacting (ideal) gas. Also as we are interested in the finite temperature effects caused by moduli dynamics, we disregard ${\cal O}(T^4)$ moduli independent term in $V_T^{1-loop}$ and consider only the $T^2$ moduli dependent one. Here $g_B$ and $g_F$'s are bosonic and fermionic degrees of freedoms respectively and $\kappa_i$'s are some ${\cal O}(1)$ constants. Finally the last term in (\ref{eq:V_T}) which is a collection of two-loop finite temperature contributions, are dominant $T^4$ corrections in large temperature limit ($T\gg \{M_{\rm soft},M_{\rm moduli},M_{\rm modulino}\}$) which involves interactions of species among themselves beyond ideal gas approximation at 2-loops level. Also the ${\cal O}(g_{MSSM}^2)$, ${\cal O}(\lambda_{\chi_{(B,S)} \gamma\gamma}^2)$ and ${\cal O}(\lambda_{\tilde{\chi_{(B,S)}}\tilde{\gamma}\gamma}^2)$ contributions appear at two loops involving MSSM particles, two loops with $\chi_{(B,S)}$ and two gauge bosons, and two loops involving the modulino $\tilde{\chi_{(B,S)}}$, the gaugino e.g. $\tilde{\gamma}$ and the photon $\gamma$ respectively. The various other two-loop corrections coming from  interaction of moduli $\chi_B, \chi_S$ and (or) their respective modulini $\tilde{\chi_{(B,S)}}$'s and other MSSM particles, the self-interactions of the moduli/modulini etc. are subdominant because of negligible couplings in large volume limit.\footnote{One can see that the moduli self couplings (\ref{eq:selfcouplings}), e.g. $\delta{{\cal L}}_{\chi_B^2 \chi_S}\sim \bigl[\frac{(ln{\cal V})^{\frac{1}{2}}}{{\cal V}^{n^s-\frac{52}{36}}\ (\sum_{\beta\in H^2_{{\bar\partial},-}(CY_3)}n^0_{\beta})^2}\bigr] {\chi_B^2 \chi_S}$ are suppressed by large genus-zero Gopakumar-Vafa invariants $n^0_{\beta}\sim \frac{{\cal V}}{10}$ \cite{Sparticles_Misra_Shukla} in the large volume limit. We assume the same to be true for other similar couplings.}

As we have earlier mentioned that in the context of supporting (MS)SM-like models with magnetized D7-brane wrapping the big $\Sigma_B$ divisor of the Swiss-Cheese Calabi Yau, we have realized order one YM couplings through the gauge kinetic function by constructing appropriate involutively odd harmonic one-form and estimating the Wilson Line moduli contribution \cite{D3_D7_Misra_Shukla}:
$
f_i = \frac{T_{\rm (MS)SM}}{4\pi} + h_i(F)S;\ \ \frac{1}{g_{MSSM}^2}\sim Re({T_B})\sim {\cal O}(1)
$
where $h_i(F)$  is the flux induced contribution and hence the inverse of gauge couplings squared is given by the real part of modified $N=1$ coordinate $T_B$ in the dilute flux approximation. Now based on the pieces of information on various couplings and open as well as closed string moduli masses along with soft-susy breaking parameters available, let us estimate the runaway behavior of effective potential after the inclusion of finite temperature corrections.

\subsection*{Estimates of runaway behavior and decompactification temperature}
Now we explore the possibility of runaway behavior of local zero temperature minimum due to possible finite temperature 1-loop and 2-loops corrections in high temperature limit ($T> \{M_{\rm soft},M_{\rm moduli},M_{\rm modulino}\}$) and estimate the decompactification temperature $T^d_{max}$ upto which the zero temperature minimum remains stable and beyond which it gets destabilized. In \cite{D3_D7_Misra_Shukla}, we have realized that in our D3/D7 Swiss-Cheese orientifold setup, gaugino mass $ M_{\tilde g}\sim m_{\frac{3}{2}}$ and other soft scalar masses $D3$-brane position moduli mass $ m_{{\cal Z}_i}\sim {\cal V}^{\frac{19}{36}}m_{\frac{3}{2}}$ and Wilson line moduli mass  $ m_{\tilde{\cal A}_1}\sim {\cal V}^{\frac{73}{72}}m_{\frac{3}{2}}$ where $ m_{\frac{3}{2}}\sim{\cal V}^{-\frac{n^s}{2} - 1} M_p$. Also the various moduli masses e.g. complex structure moduli and axion-dilaton masses $M_{c.s+\tau}\sim \frac{1}{\cal V}\ M_p$ while closed string canonical K\"{a}hler moduli masses  are $M_{\chi_B}\sim {{\cal V}^{-\frac{n^s+1}{2}}}\  M_p; \ M_{\chi_S}\sim {{\cal V}^{-\frac{n^s}{2}}}\ M_p$. As argued in \cite{CicoliFTLVS}, we assume that the various modulini masses and couplings are of the same order as their respective moduli masses and couplings to matter fields. Now assuming D3-instanton number $n^s=2$,  and using various masses discussed above along with moduli-visible sector couplings, we arrive at the following dominant moduli dependent contribution due to loop corrections is:
\be
\label{012loopV}
V_{T=0}\sim {\cal V}^{-\frac{53}{18}}\ M_p^4;\ \ V^{1-loop}_T \sim \frac{T^2 M_p^2}{24 {\cal V}^2}; \ \ V^{2-loop}_T \sim T^4
\ee
As the effective field theory description are valid up to energies lower than the string scale $M_s$, we compare the finite temperature corrections to zero-temperature effective scalar potential in the temperature regime $M_s\ge T> \{M_{\rm soft},M_{\rm moduli},M_{\rm modulino}\}$ and find the following estimates on temperature for destabilizing the zero-temperature local minimum of scalar potential,
\bea
& & V^{1-loop}_T \le V_{T=0} \Longrightarrow T \le \sqrt{\frac{24}{{\cal V}^{\frac{17}{18}}}} \ M_p \sim M_s\nonumber\\
& & V^{2-loop}_T \le V_{T=0} \Longrightarrow T \le {{\cal V}^{-\frac{53}{72}}} \ M_p \sim {{\cal V}^{-\frac{17}{72}}} \ M_s \nonumber\\
\eea
The above estimates implies that the local minimum of zero-temperature effective scalar potential is stable against finite temperature effects upto $T^d_{max}\sim {{\cal V}^{-\frac{53}{72}}} \ M_p \sim (10^{13}-10^{14}) {\rm GeV}$ (for Calabi Yau volume ${\cal V}\sim 10^6$) and the finite temperature effects cause thermal decompactification above this $T^d_{max}$. Next we observe that below this decompactification limit $T^d_{max}\sim {{\cal V}^{-\frac{53}{72}}} \ M_p$, the 2-loop contribution is dominant over 1-loop contribution which seems to be violation of perturbative description of thermal corrections, however this is not the case as we have disregarded the $T^4$ term of $V^{1-loop}_T$, the same being independent of moduli masses. Also it is interesting to notice that although the moduli/modulino interactions with gauge bosons and gauginos are sufficient for thermalizing them, however the same can cause the thermal corrections which are not enough to destabilize the local minimum of zero-temperature $N=1$ effective scalar potential below $T^d_{max}\sim (10^{13}-10^{14}) {\rm GeV}$. Further as one notice that $T^d_{max}$ is nearly the same as string scale, therefore one can argue that the local minimum of zero-temperature effective scalar potential is stable against any finite temperature corrections in the regime of validity of effective Supergravity.

\section{Conclusions and Discussions}
Lets provide a few arguments in support of the fact that moduli in our setup do not cause any cosmological moduli problem before we summarize the results of the paper.

The late decay of moduli after inflation results in the dominated energy density contributions from these moduli and these effects can jeopardize the successful predictions of  nucleosynthesis and other post inflation structure formation processes. For generic moduli with masses less than TeV, there is possibility of ``cosmological Moduli problem" which is usual for light scalar fields which couple gravitationally (see for example big divisor volume modulus with mass $\sim MeV$ of \cite{AstroLVS}). In our D3/D7 Swiss-Cheese LVS orientifold setup, we find that for Calabi Yau volume ${\cal V}\sim 10^6$ and D3-instanton number $n^s=2$, we have canonical moduli masses $M_{\chi_B}\sim 10^6 {\rm TeV}$ and $M_{\chi_S}\sim 10^9{\rm TeV}$ implying that both of the moduli are heavier than gravitino ($m_{\frac{3}{2}}\sim {\cal V}^{-\frac{n^s}{2}-1}{\rm M_p}\sim10^3{\rm TeV}$). Also as the $N=1$ zero-temperature scalar potential for complex structure moduli and axion dilaton modulous is positive semi-definite and scales as $V_{\rm c.s.+\tau}\sim \frac{1}{{\cal V}^2}$ in large volume limit \cite{AstroLVS}, which is larger as compared to K\"{a}hler moduli potential (which scales as $\sim\frac{1}{{\cal V}^3}$ in large volume limits without D3/D7 inclusions). This implies that these complex structure and axion-dilaton moduli are trapped towards minimum of potential at very early times and are not producing problematic oscillations. Further we have also observed in \cite{D3_D7_Misra_Shukla}, that open string moduli masses;  the mobile spacetime filling D3-brane position moduli ${\cal Z}_i$'s and Wilson line moduli ${\cal A}_I$'s masses are $M_{{\cal Z}_i}\sim {\cal V}^{\frac{19}{36}} m_{3/2}\sim 10^6 {\rm TeV}$ and $M_{{\cal A}_I}\sim {\cal V}^{\frac{73}{72}} m_{3/2}\sim 10^9 {\rm TeV}$ respectively and thus we find that all open and closed string moduli in our setup are heavier enough and hence naturally free from any ``Cosmological Moduli Problem". However such huge moduli masses implies that moduli are less stable and decays very fast to other particles via several allowed channels opening up the possibility of overproduction of gravitino. In our setup, we find that both closed string moduli are heavy enough to be free from cosmological moduli problem and one of those $\chi_B$ does not cause gravitino overproduction due to its gravitino-moduli coupling being volume suppressed. However, for "small" divisor modulus $\chi_S$ we find order one gravitino production branching ratio, which needs some resolution like \cite{Thermalinflation1}. Moreover the reheating temperature after moduli thermalization comes out to be very large, $T^{\chi_B}_{R}\sim \sqrt{M_p \Gamma_{\chi_B}}\sim 10^5 {\rm GeV}$ and $T^{\chi_S}_{R}\sim \sqrt{M_p \Gamma_{\chi_S}}\sim 10^{10} {\rm GeV}$ in our setup which implies the commencement of a Hot Big Bang at a relatively early stage facilitating the necessary initial conditions for a period of thermal inflation. All these attractive features of our large volume Swiss-Cheese setup put the same to be more promising for realistic model building in string cosmology, as the same setup realizes dS minima without need of any uplifting, axionic slow-roll inflation without any $\eta$-problem and required $60$ number of e-foldings along with non-trivial $f_{NL}$ as well as tensor-to-scalar ratio $r$ with loss of scale invariance $|n_R-1|=0.014$ lying well within the experimental bounds. To have the aforementioned issues (all of which being among the most challenging ones in string cosmology) addressed in a single string theoretic setup is really exciting.

To summarize in a nutshell, we present a systematic study of moduli dynamics and explore on various processes for thermalizing the moduli in large volume scenarios in the context of a type IIB compactification on an orientifold of a Swiss-Cheese Calabi Yau in the presence of a single spacetime filling mobile D3-brane, stacks of D7-brane wrapping the ``big" divisor $\Sigma_B$ along with the internal fluxes. Unlike KKLT(-like) models, we argue that in large volume scenarios, the couplings of closed string moduli are very significant, making the interactions to be string-mass suppressed instead of Planck suppressed and various interaction processes are allowed for providing the way for the moduli to get thermalized. Also, we find that closed string divisor volume moduli are heavy enough and hence do not create any cosmological problem like spoiling the postinflationary structure formation process, the nucleosynthesis. We also investigate the possibilities of destabilizing the zero-temperature de-Sitter solutions after including thermal corrections up to two-loops and similar to the observations of finite temperature effects in O'KKLT \cite{OKKLT}, we conclude with the de-Sitter solutions to be stable against any finite temperature corrections within the regime of validity of the supergravity.

\section*{Acknowledgements}

The work of PS is supported by a Senior Research Fellowship from the CSIR, India. PS would like to thank Aalok Misra for his encouragement, clarifications and discussions throughout the work whenever it was needed.

\appendix
\section{Derivatives of K\"{a}hler potential}
For this article to make self contained, we provide some intermediate steps of first and second derivatives of closed string moduli (the divisor volumes $\tau_B$ and $\tau_S$) as well as the open string moduli, namely the spacetime filling mobile D3-brane position moduli $z_i$'s (identified with two Higges) and the single Wilson line modulus ${\cal A}_I$ (identified with any of squarks/sleptons of the first two families) as below \cite{D3_D7_Misra_Shukla}.
\begin{itemize}
\item{\begin{equation}
\label{eq:singleder_sigma}
\frac{\partial K}{\partial\tau^\alpha}=-\frac{2}{\cal Y}\left[\frac{3a}{2}(2\tau_\alpha + \mu_3l^2{\cal V}^{\frac{1}{18}} + ... -\gamma K_{\rm geom})\right]^{\frac{1}{2}}.
\end{equation}}

\item
\begin{eqnarray}
\label{eq:singleder_z}
& & \frac{\partial K}{\partial z_i}=
-\frac{2}{\cal Y}\Biggl[\frac{3a}{2}\left(2\tau_B + \mu_3l^2{\cal V}^{\frac{1}{18}} + ... -\gamma K_{\rm geom}\right)^{\frac{1}{2}}\Biggl\{3i\mu_3\l^2(\omega_B)_{i{\bar j}}{\bar z}^{\bar j}+\frac{3}{4}\mu_3l^2\Bigl((\omega_B)_{i{\bar j}}{\bar z}^{\tilde{a}}({\cal P}_{\tilde{a}})^{\bar j}_lz^l \nonumber\\
& & + (\omega_B)_{l{\bar j}}z^l{\bar z}^{\tilde{a}}({\cal P})^{\bar j}_i\Bigr) -\gamma(ln {\cal V})^{-\frac{7}{12}}{\cal V}^{\frac{29}{36}}\Biggr\}
- \frac{3a}{2}\left(2\tau_S + \mu_3l^2{\cal V}^{\frac{1}{18}} + ... -\gamma K_{\rm geom}\right)^{\frac{1}{2}}\Biggl\{3i\mu_3\l^2(\omega_S)_{i{\bar j}}{\bar z}^{\bar j}\nonumber\\
& & +\frac{3}{4}\mu_3l^2\left((\omega_S)_{i{\bar j}}{\bar z}^{\tilde{a}}({\cal P}_{\tilde{a}})^{\bar j}_lz^l + (\omega_S)_{l{\bar j}}z^l{\bar z}^{\tilde{a}}({\cal P})^{\bar j}_i\right) -\gamma(ln {\cal V})^{-\frac{7}{12}}{\cal V}^{\frac{29}{36}}\Biggr\} \Biggr]
\end{eqnarray}

\item
\begin{equation}
\label{eq:singleder_a}
\frac{\partial K}{\partial {\cal A}^I}=-\frac{2}{{\cal Y}}\left[\left(2\tau_B + \mu_3l^2{\cal V}^{\frac{1}{18}} + ... -\gamma K_{\rm geom}\right)^{\frac{1}{2}}.6i\kappa^4\mu_7(C_B)^{I{\bar K}}{\bar {\cal A}}_{\bar K} \right]
\end{equation}

\item{ \begin{eqnarray}
\label{eq:double_sigmasigma}
& & \frac{\partial^2K}{{\bar\partial}{\bar\tau}^\alpha\partial\tau^\alpha}=\frac{2}{{\cal Y}^2}
\left[\frac{3a}{2}\sqrt{2\tau_\alpha + \mu_3l^2{\cal V}^{\frac{1}{18}} + ... -\gamma K_{\rm geom}}\right]^2 -\frac{3a}{2{\cal Y}}\frac{1}{\sqrt{2\tau_\alpha + \mu_3l^2{\cal V}^{\frac{1}{18}} + ... -\gamma K_{\rm geom}}}\nonumber\\
\end{eqnarray}}
\item {
\bea
\label{eq:double_sigmaBsigmaS}
& & \hskip-0.3in\frac{\partial^2K}{{\bar\partial}{\bar\tau}^S\partial\tau^B}=\frac{2}{{\cal Y}^2}
\left[\frac{3a}{2}\sqrt{2\tau_S + \mu_3l^2{\cal V}^{\frac{1}{18}} + ... -\gamma K_{\rm geom}}\right]
\left[\frac{3a}{2}\sqrt{2\tau_B + \mu_3l^2{\cal V}^{\frac{1}{18}} + ... -\gamma K_{\rm geom}}\right]\nonumber\\
\eea}

\item
\begin{eqnarray}
\label{eq:Double_zz}
& & \hskip -1cm \frac{\partial^2K}{\partial z_i{\bar\partial} {\bar z}_{\bar j}}=
\frac{2}{{\cal Y}^2}\Biggl[\frac{3a}{2}\left(2\tau_B + \mu_3l^2{\cal V}^{\frac{1}{18}} + ... -\gamma K_{\rm geom}\right)^{\frac{1}{2}}\Biggl\{3i\mu_3\l^2
(\omega_B)_{i{\bar k}}{\bar z}^{\bar k}+\frac{3}{4}\mu_3l^2\Bigl((\omega_B)_{i{\bar k}}{\bar z}^{\tilde{a}}
({\cal P}_{\tilde{a}})^{\bar k}_lz^l\nonumber\\
& & \hskip -1cm + (\omega_B)_{l{\bar k}}z^l{\bar z}^{\tilde{a}}({\cal P})^{\bar k}_i\Bigr) -\gamma(ln {\cal V})^{-\frac{7}{12}}{\cal V}^{\frac{29}{36}}\Biggr\}
- \frac{3a}{2}\left(2\tau_S + \mu_3l^2{\cal V}^{\frac{1}{18}} + ... -\gamma K_{\rm geom}\right)^{\frac{1}{2}}\Biggl\{3i\mu_3\l^2(\omega_S)_{i{\bar k}}
{\bar z}^{\bar k}\nonumber\\
& & \hskip -1cm +\frac{3}{4}\mu_3l^2\left((\omega_S)_{i{\bar k}}{\bar z}^{\tilde{a}}({\cal P}_{\tilde{a}})^{\bar j}_lz^l
+ (\omega_S)_{l{\bar k}}z^l{\bar z}^{\tilde{a}}({\cal P})^{\bar k}_i\right) -\gamma(ln {\cal V})^{-\frac{7}{12}}{\cal V}^{\frac{29}{36}}\Biggr\} \Biggr] \times\Biggl[\frac{3a}{2}\Bigl(2\tau_B + \mu_3l^2{\cal V}^{\frac{1}{18}} + ...\nonumber\\
& & \hskip -1cm -\gamma K_{\rm geom}\Bigr)^{\frac{1}{2}} \Biggl\{-3i\mu_3\l^2
(\omega_B)_{k{\bar j}}z^k-\frac{3}{4}\mu_3l^2\left((\omega_B)_{k{\bar j}}z^{\tilde{a}}
({\cal P}_{\tilde{a}})^k_{\bar l}{\bar z}^l + (\omega_B)_{{\bar l}k}{\bar z}^{\bar l}z^{\tilde{a}}
({\cal P})^k_{\bar i}\right)-\gamma(ln {\cal V})^{-\frac{7}{12}}{\cal V}^{\frac{29}{36}}\Biggr\}\nonumber\\
& & \hskip -1cm - \frac{3a}{2}\left(2\tau_S + \mu_3l^2{\cal V}^{\frac{1}{18}} + ... -\gamma K_{\rm geom}\right)^{\frac{1}{2}}\Biggl\{-3i\mu_3\l^2(\omega_S)_{k{\bar j}}
z^k-\frac{3}{4}\mu_3l^2\left((\omega_S)_{k{\bar j}}z^{\tilde{a}}({\cal P}_{\tilde{a}})^j_{\bar l}{\bar z}^{\bar l}
+ (\omega_S)_{k{\bar l}}{\bar z}^{\bar l}z^{\tilde{a}}({\cal P})^k_{\bar i}\right)\nonumber\\
& & \hskip -1cm -\gamma(ln {\cal V})^{-\frac{7}{12}}{\cal V}^{\frac{29}{36}}\Biggr\} \Biggr]
-\frac{2}{{\cal Y}}\Biggl[\frac{3a}{2}\left(2\tau_B + \mu_3l^2{\cal V}^{\frac{1}{18}} + ... -\gamma K_{\rm geom}\right)^{\frac{1}{2}}\left\{
3i\mu_3l^2(\omega_B)_{i{\bar j}}-\gamma\left(ln {\cal V}\right)^{-\frac{7}{12}}{\cal V}^{\frac{5}{18}}\right\}\Biggr]\nonumber\\
& & \hskip -1cm \sim \frac{{\cal V}^{1/36}}{{\sum_{\beta\in H^2_{{\bar\partial},-}(CY_3)}n^0_{\beta}}}
\end{eqnarray}

\item
\begin{eqnarray}
\label{eq:Double_aa}
& & \hskip -1cm \frac{\partial^2K}{\partial{{\cal A}^I}{\bar\partial}{\bar{\cal A}_J}}=-\frac{2}{{\cal Y}}\Biggl[\frac{3a}{4}\frac{(6i\kappa_4^2\mu_7)^2
(c_B)^{I{\bar K}}{\cal A}_{\bar K}(c_B)^{L{\bar J}}{\cal A}_L}{\left(2\tau_B + \mu_3l^2{\cal V}^{\frac{1}{18}} + ... -\gamma K_{\rm geom}\right)^{\frac{1}{2}}}
+ \frac{3a}{2}\left(2\tau_B + \mu_3l^2{\cal V}^{\frac{1}{18}} + ... -\gamma K_{\rm geom}\right)^{\frac{1}{2}}\nonumber\\
& &  \times 6i\kappa^2\mu_7(c_B)^{I{\bar J}}\Biggr]
 +\frac{2}{{\cal Y}^2}\Biggl[\frac{3a}{2}\left(2\tau_B + \mu_3l^2{\cal V}^{\frac{1}{18}} + ... -\gamma K_{\rm geom}\right)^{\frac{1}{2}}.6i\kappa^2\mu_7(c_B)^{I{\bar K}}{\bar a}_{\bar K}
\Biggr] \nonumber\\
& & \times\Biggl[\frac{3a}{2}\left(2\tau_B + \mu_3l^2{\cal V}^{\frac{1}{18}} + ... -\gamma K_{\rm geom}\right)^{\frac{1}{2}}.6i\kappa^2\mu_7(c_B)^{L{\bar J}}
{\cal A}_L \Biggr]\nonumber\\
& & \sim \frac{{\cal V}^{65/36}}{{\sum_{\beta\in H^2_{{\bar\partial},-}(CY_3)}n^0_{\beta}}}
\end{eqnarray}

\end{itemize}


\end{document}